
\magnification=1200
\def\oneandhalfspace{\baselineskip 15pt}
\def\ref{\par\noindent\hangindent 20pt}
\oneandhalfspace

\null\vskip 2cm
\centerline{\bf Perturbation growth and cosmic microwave background
anisotropies}
\centerline{\bf in the string-like matter dominated universe}
\vfill
\centerline{J Stelmach}
\bigskip
\centerline{Institute of Physics, University of Szczecin, Wielkopolska 15,}
\centerline{70-451 Szczecin, Poland}
\vfill
{\bf Abstract}. The growth of small perturbations and the temperature
correlation function of cosmic background radiation induced by the
perturbations in the isotropic and homogeneous universe with arbitrary spatial
curvature, filled with nonrelativistic and so-called string-like matter is
calculated. It is shown that both magnitudes are insensitive to the eventual
existence of the string-like matter in the universe. The widespread opinion
that the string-like matter dominated universe suppresses the growth of
perturbation is verified.
\vfill\supereject
\parindent=0pt\par
{\bf 1. Introduction}
\medskip\par
The discrepancy between observationally determined value of the energy density
parameter $\Omega_{\rm obs}=0.1\div 0.3$ and the theoretically favoured one
$\Omega \approx 1$ (inflationary scenario) suggests that the Universe can be
filled with some kind of matter which is distributed more smoothly than the
baryonic matter. The smooth component of the Universe must be nonbaryonic but
it can be of nonrelatyvistic (e.g. massive neutrinos), relativistic, or more
exotic (e.g. $\Lambda$-term) type. The problem appears in the case of
nonrelativistic and relativistic matter dominated universe since in both cases
the age of the universe becomes too short (shorter than the age of the oldest
stars and globular clusters). This is the main reason why cosmological models
with the $\Lambda$-term are in recent years so seriously taken into account.
However, the constant $\Lambda$-term is not the only possibility which resolves
the so called ``$\Omega$-problem'' giving, at the same time, sufficiently large
age of the universe. Some people consider the $\Lambda$-term varying in time
according to the law $\Lambda\propto t^{-2}$ [1,2] or $\Lambda\propto R^{-2}$
[3,4] (in some cases both behaviours coincide). Such time-dependent
$\Lambda$-term allows to explain its present extremly small value comparing
with the natural value close to the Planck epoch. We note that the global
texture in a closed universe [5] and a network of cosmic strings [6,7] obeys
the same law of variation ($\rho\propto R^{-2}$). We call the exotic form of
matter scaling according to this law -- string-like matter. In the present
paper we discuss perturbation growth and anisotropies of the cosmic microwave
background (CMB) in the string-like matter dominated universe.
\parindent=16pt\par
The influence of, what we call, string-like matter on the evolution of the
universe and on astrophysical formulae was examined by several authors [8-13].
Perturbation growth in such a model was also discussed [14]. It is well known
that the perturbations cannot grow in the curvature dominated universe. Since
the energy density of the string-like matter scales with the expansion in the
same way as the curvature term one asserts that the perturbations cannot grow
in the string-like matter dominated universe either. We find this generally
accepted statement not quite correct and one of the purposes of the paper is to
clear up this point. The second purpose of the paper is to check whether the
eventual existence of the string-like matter in the universe may have an impact
on the large-angular-scale anisotropies of the CMB. In the previous paper [12]
we
showed that the string-like matter enlarges the angle at which we observe the
anisotropies of the CMB. One can say that the string-like matter acts as a sort
of magnifying glass apparently increasing the angular size of anisotropy. The
question we address in the present paper is whether the string-like matter
influes the amplitude of the anisotropy.
\par In the next section we calculate the growth of small perturbations in the
homogeneous and isotropic universe with arbitrary spatial curvature filled with
nonrelativistic and string-like matter. Section 3 is devoted to evaluation of
the temperature correlation function of cosmic microwave background radiation
induced by the perturbations in the model under consideration. We realize that
neither the growth of perturbations nor the temperature correlation function of
the CMB is sensitive to the eventual existence of the string-like matter in the
universe.
\bigskip\parindent=0pt\par
{\bf 2. Perturbation growth in the universe filled with nonrelativistic and
string-} \item {} {\bf -like matter}
\medskip\par
If we assume that the components of the universe do not interact with each
other the Friedman equation can be written in the form
$${1\over R^2}\left(dR\over dt\right)^2={C_m\over R^3}+{C_s\over R^2}-{k\over
R^2}, \eqno(2.1)$$
where $k=0,\pm 1$ is a curvature index and $C_m$ and $C_s$ are some constants
which can be expressed in terms of astronomical parameters [10]
$$\eqalignno{C_m&={H_0}^2\Omega_{m0},&(2.2a)\cr
C_s&={H_0}^2\Omega_{s0},&(2.2b)\cr
k&={H_0}^2(\Omega_{m0}+\Omega_{s0}-1).&(2.2c)\cr}$$
$H_0$ is a present value of the Hubble constant and $\Omega_{m0}$ and
$\Omega_{s0}$ are energy density parameters of nonrelativistic and string-like
matter respectively. We use scale factor normalization such that at the present
epoch $R(t_0)=1.$ With the help of expressions (2.2) the Friedman equation
reads
$$\left(dR\over dt\right)^2={\Omega_{m0}{H_0}^2\over R}-{H_0}^2(\Omega_{m0}-1).
\eqno(2.3)$$
Note that the string-like matter energy density parameter $\Omega_{s0}$ does
not explicitly appear in the above equation. It means that the string-like
matter does not affect the dynamics of the universe. It follows from the
relation (2.2c) that $\Omega_{s0}$ does affect only the curvature. This well
known property will be crucial for our further considerations.
\parindent=16pt\par
Similar form of the curvature term and the string-like matter term in the
Friedman
equation suggests introducing, what we call, ``effective curvature term''
$k'/R^2$, where
$$k'\equiv k-C_s={H_0}^2(1-\Omega_{m0}).\eqno(2.4)$$
This term affects directly the dynamics of the universe. The curvature
term and the string-like matter term separately do not.
\par Let us assume that the string-like matter is distributed smoothly in the
universe (at a large scale) i.e. is not affected by a local perturbation of
nonrelativistic matter $\delta\rho(t,\vec x)=\rho(t,\vec x)-\bar\rho(t).$
Then the time evolution of the density contrast $\delta\equiv
\delta\rho/\bar\rho$
in the linear regime ($\delta\ll 1$) is described by the equation
$$\ddot\delta+2{\dot R\over R}\dot\delta-{3\over 2}{\Omega_{m0}{H_0}^2\over
R^3}\delta=0.\eqno(2.5)$$
Since the string-like matter forms a smoth background it does not explicitly
enter the above equation [15]. Its solution (growing mode) written down as a
function of redshift, up to some multiplicative constant reads [16]
$$\eqalignno{\delta(z)&={1+2\Omega_{m0}+3\Omega_{m0}z\over
(1-\Omega_{m0})^2}\cr
&-{3\over 2}{\Omega_{m0}(1+z)\over (1-\Omega_{m0})^{5/2}}\sqrt{1+\Omega_{m0}z}
\ln{\sqrt{1+\Omega_{m0}z}+\sqrt{1-\Omega_{m0}}\over \sqrt{1+\Omega_{m0}z}-
\sqrt{1-\Omega_{m0}}}.&(2.6)\cr}$$
Of course this is exactly the same solution as in the case without the
string-like matter as it could have been expected looking at the form of the
equation (2.5). From the solution it follows that for fixed $z$ the growth of
perturbation in the linear regime is totally determined by the nonrelativistic
matter energy density parameter $\Omega_{m0}$. The string-like matter energy
density parameter $\Omega_{s0}$ has nothing to do with the growth $\delta(z)$.
For fixed $\Omega_{m0}$ the growth is the same independently of whether
$\Omega_{s0}=0$ or $\Omega_{s0}\gg\Omega_{m0}.$ The same conclusion can be
deduced from the form of the equation (2.1) or (2.3). In the case without the
string-like matter perturbations cease to grow if only the curvature term
starts to dominate, i.e. for $\vert k\vert/R^2>C_m/R^3.$ But in the case under
consideration the curvature term is replaced by the ``effective curvature
term'' $k'/R^2$, which is a difference of the curvature term and the
string-like matter term. In this difference $\Omega_{s0}$ cancels. As a result
the
redshift at which the perturbation ceases to grow is independent of the value
of $\Omega_{s0}$
$$z_c={1-\Omega_{m0}\over \Omega_{m0}}-1.\eqno(2.7)$$
\bigskip\parindent=0pt\par
{\bf 3. Temperature correlation function in the model with nonrelativistic and}
\item {} {\bf string-like matter}
\medskip
The perturbation growth calculated in the previous section by local change of
gravitational potential induces anisotropy in the cosmic microwave background
radiation. This anisotropy can be calculated in several ways. In the present
paper we are using a method proposed by G\'orski {\it et al} [17]. In this
method an
expression for the temperature correlation function $C(\theta)$ involves a
double integral along the lines of sight, separated by the angle $\theta$, to
the last scattering surface over the second derivative of the growth rate
factor for adiabatic density fluctuation in nonrelativistic matter combined
with simple integral transforms of the power spectrum of inhomogeneity
$$C(\theta)=\int\int\limits_0^{\eta_0-\eta_e}ds_1ds_2\ddot\delta(\eta_e+s_1)
\ddot\delta(\eta_e+s_2)\left[\psi_\parallel(y)\cos\theta+\psi_-(y){y_1y_2\over
y^2}\sin^2\theta\right].\eqno(3.1)$$
$\ddot\delta(\eta)$ is the second derivative of the perturbation $\delta$ with
respect to the conformal time $\eta$ defined $d\eta=H_0dt/R(t)$. Using the
evolution equation (2.5) written down in a form with $\eta$-variable
$$\ddot\delta(\eta)+{\dot R(\eta)\over R(\eta)}\dot\delta(\eta)-{3\over 2}
\Omega_{m0}{\delta(\eta)\over R(\eta)}=0\eqno(3.2)$$
$\ddot\delta(\eta)$ can be expressed as
$$\ddot \delta(\eta)=(1-\Omega_{m0})\left({9\over
\cosh\eta-1}+2\right)\delta(\eta)-1,\eqno(3.3)$$
where the density contrast $\delta$ as a function of $\eta$ is
$$\eqalignno{\delta(\eta)&={1\over {H_0}^2(1-\Omega_{m0})}\Biggl[1+{6\over
\cosh\eta-1}+
{3\over \cosh\eta-1}\left({2\over \cosh\eta-1}+1\right)^{1/2}\cr
&\times\ln{\left({2\over
\cosh\eta-1}+1\right)^{1/2}-1\over \left({2\over \cosh\eta-1}+1\right)^{1/2}+1}
\Biggr].&(3.4)\cr}$$
In order to find $\delta(\eta)$ in the above form we integrated the Friedman
equation
$$\left(dR\over d\eta\right)=\Omega_{m0}R+(1-\Omega_{m0})R^2\eqno(3.5)$$
and the solution is
$$R(\eta)={\Omega_{m0}\over 2(1-\Omega_{m0})}(\cosh\eta-1).\eqno(3.6)$$
$\eta_0$ in the upper limit of the integration in (3.1) is the value of the
time parameter corresponding to the present moment and can be calculated from
(3.6) by putting $R(\eta_0)=1.$
$$\eta_0={\rm arcosh}{2-\Omega_{m0}\over\Omega_{m0}}.\eqno(3.7a)$$
$\eta_e$ corresponds to the last scattering surface and can be found similarly
by putting $R(\eta_e)=1/(z+1)$, where $z$ is the redshift of the surface
($z\approx 1100).$
$$\eta_e={\rm
arcosh}\left[{2(1-\Omega_{m0})\over\Omega_{m0}(z+1)}+1\right].\eqno(3.7b)$$
$y_1, y_2$ and $y$ are defined
$$\eqalignno{y_1&=c(\eta_0-\eta_e-s_1),&(3.8a)\cr
y_2&=c(\eta_0-\eta_e-s_2),&(3.8b)\cr
y^2&={y_1}^2+{y_2}^2-2y_1y_2\cos\theta.&(3.8c)}$$
The functions $\psi_\parallel(y)$ and $\psi_-(y)$ are given by
$$\psi_\parallel(y)={1\over 2\pi^2c^2}\int_0^\infty dkP(k){dj_1(ky)\over dky},
\eqno(3.9a)$$
$$\psi_-(y)={1\over2\pi^2c^2}\int_0^\infty dkP(k)j_2(ky),\eqno(3.9b)$$
where $j_1$ and $j_2$ are spherical Bessel functions
$$\eqalignno{j_1(x)&={\sin x-x\cos x\over x^2},&(3.10a)\cr
j_2(x)&={(3-x^2)\sin x-3x\cos x\over x^3}.&(3.10b)}$$
Finally $P(k)$ is the perturbation power spectrum which is choosen in a
standard form [18,19]
$$P(k)={Ak^n\over (1+\alpha k+\beta k^{3/2}+\gamma k^2)^2},\eqno(3.11)$$
where
$$\eqalignno{\alpha &=170(\Omega_{m0}h)^{-1}{\rm km/s},&(3.12a)\cr
\beta &=9\times 10^3(\Omega_{mo}h)^{-3/2}({\rm km/s})^{3/2},&(3.12b)\cr
\gamma &=10^4(\Omega_{m0}h)^{-2}({\rm km/s})^2.&(3.12c)}$$
$h$ is the Hubble constant in units of 100 ${\rm km s}^{-1}{\rm Mpc}^{-1}.$ The
discrepancy between the numerical values of the constanys $\alpha, \beta,
\gamma$
given in original papers and the values given above follows from the different
choice of
units for $k$. Usually one chooses $[k]={\rm Mpc}^{-1}.$ Following G\'orski
{\it et
al.} we choose $[k]=({\rm km/s})^{-1}$. The latter units are obtained from the
former
ones by dividing them by the factor $H_0=100~h {\rm km/sMpc}$. $n$ is the
spectral
index usually taken as $n\approx 1$ (Harrison-Zel'dovich spectrum). The
constant $A$ can be found from the
normalization condition [17]
$$1={\delta^2(\eta_0)\over 2\pi^2}\int_0^\infty dk k^2P(k)\left[3j_1(800k)\over
800k\right]^2.\eqno(3.13)$$
$\delta(\eta_0)$ can be obtained from (2.6) by putting $z=0$.
\parindent=16pt\par
In order to find the temperature correlation function $C(\theta)$ in the
model, the expressions for $\eta_0, \eta_e, \delta(\eta)$ and $\ddot
\delta(\eta)$ should be now substituted into the formula (3.1). Numerical
determination of the function $C(\theta)$ is long-lasting because the
multiple integration has to be carried out. But even without integration some
conclusions may be deduced. It seems that the crucial one is that the
temperature correlation function $C(\theta)$, or in the consequence, the
anisotropy amplitude $\delta T/T(\theta)=[2(C(0)-C(\theta))]^{1/2}$ does not
depend on the amount of string-like matter in the universe. The amplitude is
the same in the model with and without the string-like matter.
\bigskip\parindent=0pt\par
{\bf 4. Conclusions}
\smallskip\par
In a widespread opinion the suppression of the perturbation growth is a most
important flaw of the string-like matter dominated universe practically
eliminating it as a model of the real universe.
\parindent=16pt\par
One can easily show that in the model with two components (nonrelativistic and
string-like matter) noninteracting with each other the string-like matter
starts to dominate at the redshift
$$z_s={\Omega_{s0}\over \Omega_{m0}}-1.\eqno(4.1)$$
Note that unless $k=0$ the value of $z_s$ has nothing to do with the value of
the redshift $z_c$ at which the ``effective curvature term'' starts to dominate
$$z_c={1-\Omega_{m0}\over \Omega_{m0}}-1.\eqno(4.2)$$
Formally it might happen that the universe is dominated by the string-like
matter and the perturbations still grow because $z_s>z_c$ (in the case when
$\Omega_{s0} +\Omega_{m0}-1>0$).
\par In the paper we showed explicitly that the perturbation growth in the
linear regime is totally independent of the amount of the string-like matter in
the universe. The growth is completely determined by the amount of
nonrelativistic matter. For example if we fix $\Omega_{m0}<1$, then for
$\Omega_{s0}<1-\Omega_{m0} ~(k=-1)$ or $\Omega_{s0}=1-\Omega_{m0} ~(k=0)$
or $\Omega_{s0}>1-\Omega_{m0} ~(k=1)$ the total perturbation growth is the
same.
One can say that the perturbation growth does not feel the existence of the
string-like matter. What distinguishes all three cases is the spatial
curvature. At the first sight it seems strange that the growth rate of the
perturbation does not depend on the curvature of the universe. This, however,
should not be surprising because the curvature term and the string-like matter
term enter the dynamical equations always together as an ``effective curvature
term'' and the curvature in the extracted form does not appear.
\par Since the calculation of the temperature correlation function of the CMB
requires practically the same equations as the calculation of the perturbation
growth the amplitude of the anisotropy of the CMB does depend neither on the
curvature nor on the $\Omega_{s0}$-parameter either. It means that the
measurement of the amplitude of the anisotropy $\delta T/T(\theta)$ does not
say anything about the eventual existence of the string-like matter in the
universe.
\par Conluding we would like to stress once more that the formation of
structure in the universe with or without the string-like matter proceeds in
the same way. There is no suppression of the perturbation growth in the
string-like matter dominated universe. Moreover such exotic matter has no
effect on the value of the anisotropy amplitude $\delta T/T(\theta)$.
\bigskip\parindent=0pt\par
{\bf References}
\ref [1] Berman M S 1991 {\it Phys. Rev.} D {\bf 43} 1075
\ref [2] Lopez J L and Nanopoulos D V 1995 preprint CERN-TH/95-6
\ref [3] \H Ozer M and  Taha M O 1987 {\it Nucl. Phys.} {\bf B287} 776
\ref [4] Abdel-Rahman A -M M 1990 {\it Gen. Rel. Grav.} {\bf 22} 655
\ref [5] Davies R L 1987 {\it Phys. Rev.} D {\bf 35} 3705
\ref [6] Vilenkin A 1984 {\it Phys. Rev. Lett.} {\bf 53} 1016
\ref [7] Kardashev N S 1986 {\it Sov. Astron.} {\bf 30} 498
\ref [8] Kibble T W B 1976 {\it J. Phys} A {\bf 9} 1387
\ref [9] Charlton J and Turner M S 1987 {\it Astrophys. J} {\bf 313} 495
\ref [10] D\c abrowski M P and  Stelmach J {\it Astron. J.} {\bf 97} 978
\ref [11] Tomita K and Watanabe K 1990 {\it Prog. Theor. Phys.} {\bf 84} 892
\ref [12] Stelmach J, D\c abrowski M P and Byrka R 1994 {\it Nucl. Phys.} B{\bf
406} 471
\ref [13] Stelmach J 1994 {\it Gen. Rel. Grav.} {\bf 26} 275
\ref [14] Turner M S 1985 {\it Phys. Rev. Lett.} {\bf 54} 252
\ref [15] Solov'eva L V and  Nurgaliev I S 1985 {\it Astron. J.} (russian)
{\bf 62}, 459
\ref [16] Zel'dovich Ya B and Novikov I D 1975 {\it Stroyenie i Evolucya
Vselennoy} (Moskva: Nauka)
\ref [17] G\'orski K M, Silk J and  Vittorio N 1992 {\it Phys. Rev. Lett.} {\bf
68} 733
\ref [18] Blumenthal G R and Primack J R 1983 {\it Fourth Workshop on
Grand Unification}, ed. Weldon H A, Langacker P and Steinhardt P J
(Boston: Birkhauser)
\ref [19] Bond J R and Efstathiou G 1984 {\it Astrophys. J} {\bf 285}, L45
\end